\pdfoutput=1 

\documentclass[11pt,twoside]{article}%%%%where rsproca is the template name

\usepackage{aspperso}
\usepackage{epsf}
\usepackage{natbib}
\usepackage{lscape}
\usepackage{graphics}
\usepackage{graphicx}
\usepackage{amsmath,amssymb}
\usepackage{mathptmx}

\begin{document}

%%%% Article title to be placed here
\title{Possible climates on terrestrial exoplanets }

\author{%%%% Author details
F. Forget and J. Leconte}

%%%%%%%%% Insert author address here
\affil{
Laboratoire de M\'et\'eorologie Dynamique, IPSL, Paris France
}

%%%% Abstract text to be placed here %%%%%%%%%%%%
\begin{abstract}

What kind of environment may exist on terrestrial planets around other stars? In
spite of the lack of direct observations, it may not be premature to speculate
on exoplanetary climates, for instance to optimize future telescopic
observations, or to assess the probability of habitable worlds. To first order,
climate primarily depends on 1) The atmospheric composition and the volatile
inventory; 2) The incident stellar flux; 3) The tidal evolution of the planetary
spin, which can notably lock a planet with a permanent night side.

The atmospheric composition and mass depends on complex processes which are
difficult to model: origins of volatile, atmospheric escape, geochemistry,
photochemistry. 
We discuss physical constraints which can help us to speculate on
the possible type of atmosphere, depending on the planet size, its final
distance for its star and the star type. Assuming that the atmosphere is known,
the possible climates can be explored using Global Climate Models analogous to
the ones developed to simulate the Earth as well as the other telluric
atmospheres in the solar system. Our experience with Mars, Titan and Venus
suggests that realistic climate simulators can be
developed by combining components like a ``dynamical core'', 
a radiative transfer
solver, a parametrisation of subgrid-scale turbulence and convection, a thermal
ground model, and a volatile phase change code. On this basis, we can aspire to
build reliable climate predictors for exoplanets. However, whatever the accuracy
of the models, predicting the actual climate regime on a specific planet will
remain challenging because climate systems are affected by strong positive
feedbacks. They can drive planets with very similar forcing and volatile
inventory to completely different states. For instance the coupling between
temperature, volatile phase changes and radiative properties results in
instabilities such as runaway glaciations and runaway greenhouse effect.

\end{abstract}
%%%%%%%%%%%%%%%%%%%%%%%%%%%

%%%%%%%%%% Insert the texts which can accomdate on firstpage in the tag "fmtext" %%%%%

%%%%%%%%%%%%%%% End of first page %%%%%%%%%%%%%%%%%%%%%

%\maketitle

\section{Introduction}

To help design future ground-based or space telescopes aiming at characterizing
the environment on terrestrial exoplanets, or to address 
scientific questions like the
probability of habitable worlds in the galaxy, 
one has to make assumptions on the possible climates and atmospheres
that may exist on terrestrial exoplanets.  
For this, speculation is unavoidable because no direct observations
of terrestrial atmospheres are available outside the solar system. 
The limited sample that we can observe here suggest that a wide diversity of
planetary environment is possible. 
Would we imagine Venus or Titan if they were
not there? 

Fortunately, observational statistics on the exoplanets themselves 
are starting to be available. The extrapolation of 
super-Earth detections
suggests that terrestrial planets should be abundant in our galaxy. 
A large fraction of the stars is likely to harbor rocky planets
\cite[]{Howa:10,Boru:11,Bonf:13,Cass:12}. 
These discoveries have also profoundly changed 
our vision of the formation, structure, and composition of 
low mass planets:  while it has been long thought, 
mostly based on the observation of our own Solar System, 
that there should be a gap between telluric planets with a 
thin, if any, secondary atmosphere and the 
so called icy giants that retained a substantial amount 
of hydrogen and helium accreted from the protoplanetary disk, 
this gap does not seem to exist in exoplanetary systems.

\begin{figure}
  \centering
\includegraphics[width=12cm,clip]{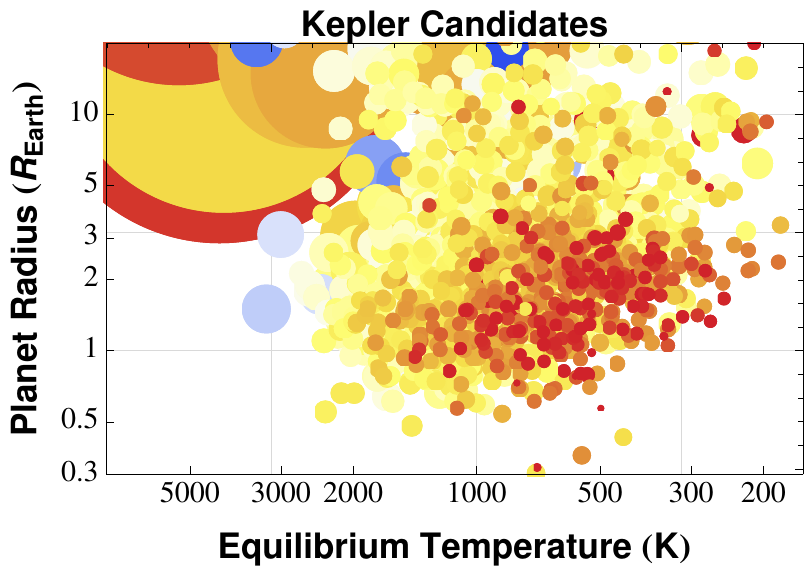}
\caption{\textit{Kepler} planet candidates in a radius-equilibrium temperature diagram. The size and the color of each dot are respectively representative of the size and color of the parent star. This diagram suggests the absence of any gap in the planet radius distribution between Earth- and Neptune-size planets.
The ``Equilibrium Temperature'' $T_e$ is obtained assuming a
planetary albedo set to zero: $T_e = (F/4\sigma)^{0.25} $, with $F$ the
mean stellar flux at the orbital distance and $\sigma$ the Stefan-Boltzmann
constant ($\sigma= 5.67~10^{-8}$ SI)
}
\label{fig:TeqR}
\end{figure}

As can be seen in Figure~\ref{fig:TeqR}, the distribution of the 
radius of planet candidates detected by the Kepler space telescope 
\cite[]{BRB13} is quite 
continuous from 0.7 up to 10 Earth radii, and particularly between 2-4 
Earth radii where the transition from Earth- to Neptune-like planets 
was thought to occur. Although these observations are still incomplete 
- especially when planets get smaller, or have a lower equilibrium 
temperature, or orbit bigger stars - they suggest that there may not be
a
clear cut distinction between low mass terrestrial planets and more 
massive planets for which the gaseous envelope represents a significant 
fraction of the bulk mass. If such a continuum exists in the bulk 
composition of low mass planets, one can also anticipate that the 
various atmospheric compositions seen in the Solar System are only 
particular outcomes of the continuum of possible atmospheres.

This raises several pending questions. What kind of atmospheres can we 
expect? Can we relate the global, measurable parameters of a planet 
(mass, radius, intensity and spectral distribution of the incoming 
stellar energy, ...) to the mass and composition of its atmosphere and 
ultimately predict a range of possible climates?

In this paper written for non specialists, we review the different
processes which may control the environment on terrestrial exoplanets, including
their habitability. In Section~\ref{sc:which}, we speculate on the possible
diversity of atmospheric composition and mass which depends on complex processes which are
not easy to model: origins of volatiles, atmospheric escape, geochemistry,
long-term photochemistry. In Section~\ref{sc:rotation}, we mention the
importance of the body rotation (period and obliquity)  and evaluate the impact
of gravitational tides on planetary spin.
If the atmosphere and the rotation are 
known, we explain in Section~\ref{sc:clim} that, 
the corresponding possible climates can be explored 
based on the likely assumption
that they are controlled by the same type of
physical processes at work on solar system bodies. 
In particular, this can be achieved using  
Global Climate Models analogous to the
ones developed to sucessfully 
simulate the Earth climate as well as Mars, Venus, Titan,
Triton, Pluto. 
However, as discussed in Section~\ref{sc:instability},
whatever the accuracy of the models, predicting the actual climate
regime on a specific planet will remain challenging because climate systems are
affected by strong positive feedbacks (e.g. 
coupling between 
radiative properties, temperatures and volatiles phase changes) 
and instabilities
which can drive planets subject to very similar
volatiles inventory and forcing to completely different states. 

%%%%%%%%%%%%%%%%%%%%%%%%%%%%%%%%%%%%%%%%%%%%%%%%%%%%%%%%%%%%%%%
%%%%%%%%%%%%%%%%%%%%%%%%%%%%%%%%%%%%%%%%%%%%%%%%%%%%%%%%%%%%%%%

\section{Which atmosphere on exoplanets?} 
\label{sc:which}
\subsection{ Origins of atmospheres } 
\label{sec:evolution_atm}

To understand the various possible types of atmospheres, one first 
needs to consider the various sources of volatiles available during the 
formation of the planet. These sources have mainly two origins: the 
nebular gas present in the protoplanetary disk during the first 1 to 
10\,Myr of the planet formation, and the volatiles (mainly H$_2$O and
CO$_2$) condensed and trapped into the planetesimals accreting on the 
nascent planet (and possibly into the comets or asteroids colliding
the planet after its formation in the so-called ``late veneer
scenario'').  The volatiles initially incorporated in the 
bulk of the mantle can be released through two major channels, 
catastrophic outgassing and release by volcanism, with very different 
timescales.

As discussed in Section~\ref{sec:atm_loss},
because atmospheric escape
is closely related to the stellar activity, it is 
strongly time dependent at early ages. The timescale on which the 
various species can be added to the atmosphere is thus critical in 
determining what is left in the matured atmosphere. Hence, we will 
discuss these three formation channels and their associated timescales 
separately.

\subsubsection{Nebular gas and protoatmospheres}

When a dense, cold molecular cloud gravitationally collapses to form a 
protostar, conservation of angular momentum forces a fraction of the 
gas to remain in an extended disk where planets can form. This gas is 
mainly composed of hydrogen and helium. The abundances of heavier 
elements are expected to be close to the stellar ones, except for some 
elements that can be trapped in condensing molecules. While these disks 
may be quickly dispersed by stellar radiation and winds (on timescales on 
the order of 3\,Myr, \citealt{Hall:03}), planetary embryos more massive than 
$\sim0.1\,M_\mathrm{Earth}$ can retain a significant mass of nebular gas, 
depending of course on the local conditions in the nebula, on the core 
mass and on the accretion luminosity \citep{IH12,Lamm:11}.

%%%%%%%%%%%%%%%%%%%%%%%%%%%%%%%%%%%%%%%%%%%%%%%%%%%%%%%
\begin{figure}
  \centering
\includegraphics[width=10cm,clip]{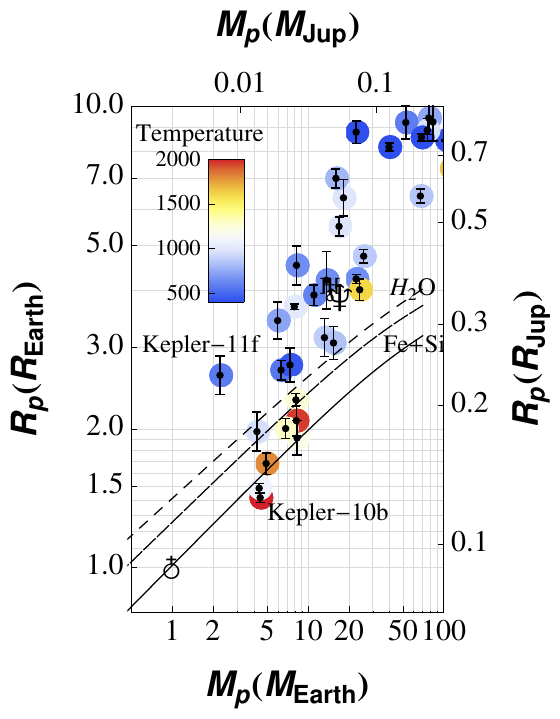}
    \caption{Mass-radius diagram of planets in the Earth to Saturn mass 
regime. The color of each dot is related to the equilibrium temperature 
of the planet (see color bar in K). Curves represent the mass radius 
relationship for an Earth like planet with a water mass fraction of 0, 
0.5 and 1 from bottom to top. Planets above the top curve must have a 
massive gaseous envelope to explain their large radii. One can see that 
in the low mass regime, hotter planets preferentially have a higher 
density that is either due to the more efficient escape or to lower gas 
accretion efficiency in hot regions of the disk.
``Temperature'' corresponds to the planetary equilibrium temperature, as
in Figure~\ref{fig:TeqR}. 
}
  \label{fig:MR}
\end{figure}
%%%%%%%%%%%%%%%%%%%%%%%%%%%%%%%%%%%%%%%%%%%%%%%%%%%%%%%

An extreme case occurs when the embryo becomes massive enough and the 
mass of the atmosphere becomes similar to the core mass. Then, the so 
called core instability can be triggered, resulting in an unstable gas 
accretion that can proceed almost until all the available gas in this 
region of the disk is fed to the planet \cite[]{Miz80,Ste82,PHB96}. This 
is the mechanism that is thought to have formed the four giant planets 
in our Solar System. The critical core mass above which the core 
instability is triggered could be as low as a few Earth masses, and a 
substantial primitive atmosphere could be accreted by much smaller planets 
\cite[]{IH12}.

This is illustrated in Figure~\ref{fig:MR} where the masses and radii 
of all the observed low mass transiting planets so far are reported. If
an 
object exhibits a radius that is bigger than the radius that would have 
a body composed entirely of water  
(water being the least dense of most abundant material 
except for H/He) of the same mass (dashed curve), this tells us that at 
least a few percents of the total mass of the planet are made of low 
density species, most likely H$_2$ and He gas. 
The fact that many objects less massive than Neptune are in this 
regime indeed confirms the possibility to accrete a large fraction of 
gas down to 2-3$\,M_\mathrm{Earth}$, the mass of Kepler-11\,f.

The fact that, in a given mass range, radii can easily vary by 
a factor of two reminds us, if need be, that the early gas accretion 
depends on many parameters that are not well understood (mass and 
dispersal time of the disk that can change from one system to another, 
location of the protoplanet, ...). The determination of the gas mass 
fraction of a given object, even knowing its current properties, is far 
from being a trivial task.

\subsubsection{Catastrophically outgassed H$_2$O/CO$_2$ atmospheres}

The other source of volatiles are the planetesimals that accrete to 
form the bulk of the planet itself. These will be the major sources of 
1) carbon compounds like CO$_2$ or possibly CH$_4$, 2) 
water, especially if they formed
beyond the ``snow line'' (the distance from the star in the nebula
where it is cold enough for water to condense into ice grains)
and, 3) to a lesser extent, N$_2$/NH$_3$ and other trace 
gases.  

In current terrestrial planet formation models, planets are usually 
formed in less than 100\,Myr. During this phase, the energy produced by 
the impacts of the planetesimals and planetary embryos is generally 
large enough to melt the upper mantle, creating a planet-wide magma 
ocean. When the accretion luminosity decreases, however, this magma 
ocean starts to solidify. Because solidification is more easily 
initiated at high pressures and molten magma is less dense than the 
solid phase, this solidification proceeds from the bottom upward 
\cite[]{Elk08}. During this phase which can last from 10$^5$\,yr to 
3~10$^6$~yr depending on the volatiles fraction, H$_2$O and CO$_2$, which 
cannot be trapped in the solid phase in large quantities, are rapidly 
outgassed.

The mass of the resulting atmosphere then depends on the composition of 
the planetesimals, and thus on their initial location as well as on the 
metallicity of the star. 

% FRANCOIS: Pour moi la partie ci-dessous
% est trop proche d'un résultat de recherche
% de Helmut pour l'inclure dans une telle review:
%On Earth, the absence of a massive hydrogen 
%.envelope resulting from the atmospheric escape of a dense steam 
%atmosphere (see \sect{sec:atm_loss}) suggests that no more than two 
%Earth oceans equivalent mass 
%(EO\,$\approx\,1.4\times10^{21}$\,kg\,$\approx 270$\,bar on Earth) of 
%water and 50-70\,bar of CO$_2$ were released that way. This corresponds 
%to rather dry planetesimals whose water mass fraction does not exceed 
%0.1\% (0.1wt\%). 

For a planet like the Earth formed at warm temperatures 
(where water ice is not stable), the available amount of volatites should be
limited because the water mass fraction in the planetesimal should be low.
It is estimated that no more than a few 
Earth oceans equivalent mass 
(EO\,$\approx\,1.4\times10^{21}$\,kg\,$\approx 270$\,bar on Earth) of 
water and 50-70\,bar of CO$_2$ were released that way (\citealt{Lamm:13} 
and reference therein).

If the planet is formed much closer to, or even 
beyond, the snow line, the water content of the planetesimals could be 
much larger (a few 10wt\%), and tens to thousands of Earth oceans of 
water could be accreted \cite[]{Elk11}. This suggests the existence of a 
vast population of planets with deep oceans (aqua-planets) or even 
whose bulk composition is dominated by water (ocean planets; 
\citealt{LSS04}). In that case, the physical state of the outer water 
layer (supercritical, steam, liquid water, ice), depends on the 
temperature that is first controlled by the cooling mantle during the 
first tens of Myr and then by the insolation received.

\subsubsection{Volcanically degassed secondary atmospheres   }

On a much longer, geological timescale, the volatiles that remained 
trapped in the mantle during the solidification can be released through 
volcanic outgassing. Along with H$_2$O and CO$_2$, this process can bring 
trace gases to the surface, such as H$_2$S, SO$_2$, CH$_4$, NH$_3$, HF, 
H$_2$, CO, and noble gases such as Ar, Xe, etc.

On Earth and Mars, there is strong evidence that this secondary 
outgassing has played a major role in shaping their present atmospheres. 
%%%%%%%%%% AJOUT FF %%%%%%%%%%%%%%%%%%%%
In particular, \cite{Tian:09} showed 
that the thermal escape (see below) induced by the 
extreme ultraviolet flux from the young sun was so 
strong  that a CO$_2$ atmosphere
could not have been maintained on Mars until about 4.1 billion years ago. 
Nevertheless, a late secondary atmosphere is thought to
have been degassed, in particular 
via the magmas that formed the large 
volcanic Tharsis province. 
\cite{Phil:01} estimated  that
the integrated equivalent of a 1.5-bar CO$_2$ atmosphere 
could have accumulated, but more realistic models have significantly 
lowered this value.
\citep{Hirs:08,Grot:11}.
Similarly, the 5\% of photochemically unstable 
methane present in the present-day 
Titan atmosphere are thought to originate from  episodic outgassing of
methane stored as clathrate hydrates within an icy shell in the interior of
Titan \cite[]{Tobi:06}.
%%%%%%%%%%%%%%%%%%%%%%%%%%%%%%%%%%%%%%%%

\subsection{ Atmospheric sinks} \label{sec:atm_loss}

While tens to thousands of bars of H/He and CO$_2$ may have been 
present in the early Earth atmosphere, they are obviously not there 
anymore (the water now being in liquid form in the oceans). This 
tells us that some processes, e.g. atmospheric escape and 
weathering/ingassing, can play the role of atmospheric sinks, and that 
these processes are powerful enough to remove completely massive 
protoatmospheres if the right conditions are met.

Considering the fact that there are three main successive delivery
mechanisms of 
different volatiles during the early stages of the planet's evolution, 
the main questions are to know when these atmospheric sinks are most 
efficient and if they can selectively deplete some species with respect 
to the others. This is what we will now discuss.

\subsubsection{Atmospheric escape}

Atmospheric molecules can 
leave the planet's attraction if they go upward with a speed exceeding 
the escape velocity \cite[]{jea25}.
However, in the lower part of the atmosphere, the gas density 
ensures a high collision rate, preventing hot particles with a 
sufficient velocity to leave the planet. As the density decreases with 
height, this assumptions breaks down when the mean free path of the 
particles becomes bigger than the scale height of the atmosphere. 
Around this level, called the exobase, stellar excitation by radiation 
and plasma flows is important, and fast enough atmospheric 
particles can actually escape.

There are several ways for the particles to reach escape
velocities , defining the various escape mechanisms that can be 
separated into two families: thermal and non-thermal escape (see 
\citealt{Lam13} for a review). 

{\bf Thermal escape} characterizes atmospheric 
escape primarily caused by the 
radiative excitation of the upper atmosphere.  
To first order, it
depends on the gravity, and on the temperature of the exobase
This temperature is not controlled by the total bolometric insolation which
heats the surface and the lower atmosphere, but by the flux of
energetic radiation and the plasma flow from the star (especially the extreme
ultraviolet which is absorbed by the upper atmosphere). It also depends on 
the ability of the atmospheric molecules to radiatively cool to space by
emitting infrared radiation; to simplify, greenhouse gases like CO$_2$ can
efficiently cool, whereas other gases like N$_2$ cannot.
Thermal escape exhibits two regimes:
 \begin{itemize} 
\item[-] \textit{Jeans escape} when the exosphere is in 
hydrostatic equilibrium, and only the particles in the high energy tail 
of the Maxwell distribution can leave the planet. 
Lighter atoms and molecules like hydrogen
and helium are more affected because they 
reach a much higher velocity at a given thermospheric temperature.
\item[-] 
\textit{hydrodynamic escape}, the so-called \textit{blow-off} regime, 
when radiative heating can only be compensated by an adiabatic 
expansion and escape of the whole exosphere. On the terrestrial planets
in our
solar system such conditions may have been reached in H- or He-rich 
thermospheres heated by the strong 
EUV flux of the young Sun
 \end{itemize} 
{\bf Non-thermal escape} result from energetic 
chemical reactions or interactions with the stellar wind 
(\textit{ion 
pick-up, plasma instabilities, cool ion outflow, polar wind}, etc.). See 
\citet{Lam13} for a more complete description of these processes which
can play a significant role on planets like modern Earth where gravity
and temperatures prevent efficient thermal escape.

{\bf Impact escape:} 
Finally, atmospheres can be lost to space because of the impacts of
comets or asteroids. 
If the gravity is low enough (thus especially for small bodies),  and if impactors are sufficiently big and fast, the hot plumes resulting from the impact can expand faster than the escape velocity and drive off the overlying air. On small planets and satellites, the efficiency of this process does not depend on the temperatures and the insolation, so that small bodies may not be able to keep an atmosphere even if atmospheric temperatures are very low.
 
While it is clearly beyond the scope of this note to go into the 
details of each of these mechanisms, it is interesting to derive an 
order of magnitude estimate for the maximum escape that can be expected 
for a given planet. This can be done by considering the hypothetical 
case of \textit{energy limited escape}. This limit is obtained by 
assuming that a given fraction $\eta$ (the heating efficiency) of the 
radiative flux available to be absorbed in the upper atmosphere is 
actually used to extract gas from the gravitational potential well of 
the planet. Because the exopheric levels are only sensitive to the very 
energetic photons in the X-EUV range (wavelength below 100~nm), 
the energy limited escape rate 
$F_\mathrm{esc}$ can be written $ F_\mathrm{esc}=\eta 
\,\frac{R_\mathrm{p}\,F_{XUV}}{G\, M_\mathrm{p}}\ \ 
\mathrm{(kg\,m^{-2}\,s^{-1})}, 
$
where $F_{XUV}$ is the 
averaged XUV flux received by the planet (i.e. divided by 4 compared to 
the flux at the substellar point), $G$ is the universal gravitational 
constant, and $R_\mathrm{p}$ and $M_\mathrm{p}$ are the planetary 
radius and mass.

Unlike the total bolometric luminosity,
the stellar luminosity in the X-EUV range $F_{XUV}$ is correlated 
with the stellar activity, which is very high at young ages and declines 
over time. Therefore, the escape rate strongly varies with time.
For example, the solar EUV flux is believed to have been 100 
times stronger than today during the first hundred million years of our 
Sun's life, to later decrease following a power law \cite[]{RGG05,Lamm:11}. 
Thermal escape is thus most relevant during the first tens to 
hundreds of Myr after the star formation, i.e. on a timescale which is 
similar to the atmosphere formation process! The implications of the 
coincidence will be discussed in Section~\ref{sec:atm_class}.

%%%%%%%%%%%%%%%%%%%%%%%%%%%%%%%%%%%%%%%%%%%%%%%%%%%%%%%%%%%%%%%%%
\begin{figure} 
  \centering
\includegraphics[width=12cm,clip]{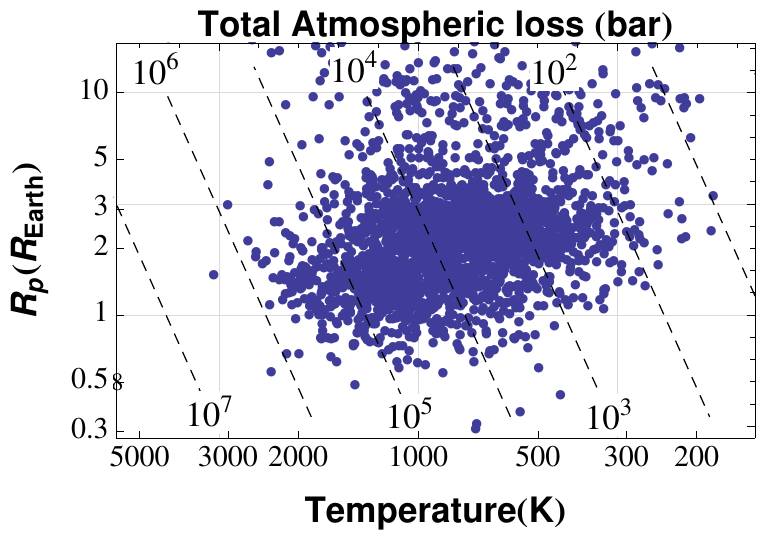} 
\caption{\textit{Kepler} planet candidates in a radius-equilibrium 
temperature diagram. Dashed lines represent contours of the average 
total amount of atmosphere that could be lost by a planet at the given 
position in the diagram in bars (atmospheric loss is integrated over 
5\,Gyr and accounts for the early active phase of the star).
``Temperature'' corresponds to the planetary equilibrium temperature, as
defined in Figure~\ref{fig:TeqR}.
} 
\vspace{-10pt} \label{fig:escape} \end{figure}
%%%%%%%%%%%%%%%%%%%%%%%%%%%%%%%%%%%%%%%%%%%%%%%%%%%%%%%%%%%%%%%%%

To give an idea of how strong atmospheric escape can be, we computed the total 
integrated atmospheric pressure that can be lost during the planet lifetime 
\begin{equation} p_\mathrm{esc}=\int 
\frac{G\,M_\mathrm{p}}{R_\mathrm{p}^2} \,F_\mathrm{esc}(t)\,\mbox{d} t
=\frac{\eta}{R_\mathrm{p}} \int F_{XUV}(t)\,\mbox{d} t, \end{equation} as a 
function of the planetary radius and for an efficiency $\eta=0.15$ 
\cite[]{MCM09,Lam13,Lamm:13}. The variation over time of the XUV to 
bolometric flux ratio (so that the results can be expressed in terms of 
the equilibrium temperature of the planet) is modeled for a solar type 
star using the parametrization of \citet{SMR11}.
The results are shown in Figure~\ref{fig:escape} (dashed labeled contours). As 
expected, atmospheres are more sensitive to escape when the planet 
receives more flux (higher equilibrium temperature) and is smaller 
(weaker gravity). Interestingly , current planet candidates (purple 
dots in Figure~\ref{fig:escape}) are expected to exhibit very different levels 
of atmospheric losses, with cold giant planets for which the effect of 
escape on the atmospheric content can almost be neglected, and highly 
irradiated Earth-like objects for which the whole atmosphere has 
probably been blown away (see Section~\ref{sec:atm_class}; \citealt{LGF11}).
%%%% revision %%%%%%%%%
Note that we did not include in our calculations the diminution of gravity
resulting from the decrease of atmospheric mass.
This can induce a positive feedback which can further accelerate atmospheric loss.
%%%%%%%%%%%%%%%%%%%%%%%

\begin{figure} 
\centering
\includegraphics[width=12cm,clip]{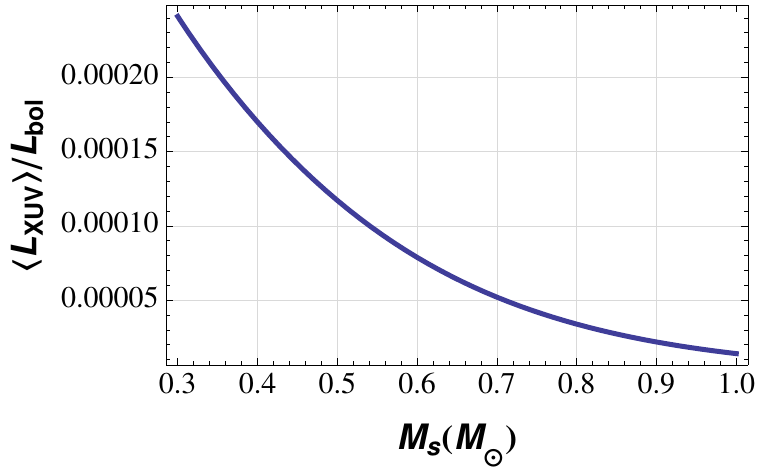} 
\caption{Ratio between the mean XUV luminosity $L_\mathrm{XUV}$ 
(between 0.1 and 100~nm and integrated over the first 5 Gyr of the star) 
and the bolometric luminosity $L_\mathrm{bol}$
of a star as a function of its mass. 
This is computed using \citet{SMR11} parametrization 
for $L_\mathrm{XUV}$. The 
greater ratio for lower mass stars stems from the longer early activity 
phase and results in a more efficient escape around smaller stars (at a 
given bolometric flux).} \label{fig:lxuv} 
\end{figure}

In Figure~\ref{fig:escape}, we assume a solar type star, but in reality,
at a given bolometric insolation, 
escape is also expected to be more intense around low mass stars as 
they emit a larger fraction of their flux in the XUV range.
This can be seen in Figure~\ref{fig:lxuv}. 
It is due to the increased duration of the active phase of lower mass stars.

%Finally, let us mention the fact that, because escape originates from 
%.the upper atmosphere, only the particles present there are 
%substantially affected. This explains why low mass atoms and molecules, 
%which are more abundant in the high levels of the atmosphere, escape 
%more easily. Hydrogen, whether it comes form the photodissociation of 
%H$_2$ or H$_2$O, is a perfect example and is generally the first gas to 
%escape, although it can drag heavier atoms along if escape occurs in 
%the hydrodynamic regime \cite[]{HPW87}.

\subsubsection{Weathering and ingassing}

\label{sc:weathering}

The atmospheric composition can also be altered by interactions with 
the surface. Indeed, if the right conditions are met, some constituents 
of the atmosphere can chemically react with the surface and get trapped 
there. In addition to this process, called weathering, the 
aforementioned surface can be burried by lava flows 
or subducted by plate tectonics, 
re-enriching the mantle in the volatiles that have been trapped; 
volatiles that will eventually be released by volcanic activity later.

An  example of such a process is the chemical weathering of 
silicate minerals 
in the presence of liquid water and CO$_2$. 
Atmospheric CO$_2$ goes into solution
in liquid water relatively easily and
the resulting fluid, carbonic acid (H$_2$CO$_3$),
reacts with the silicate minerals of the crust to
weather the rock and release calcium, magnesium,
and iron ions into the water. These ions
can promote the precipitation of solid carbonates,
(Ca,Mg,Fe)CO$_3$.  To simplify, the following net reaction 

%CaSiO$_3$(s)+CO$_2$(g) $\rightleftarrows$ CaCO$_3$(s)+SiO$_2$(s), 

\begin{align}
 \mathrm{CaSiO_3(s)+CO_2(g)\rightleftarrows CaCO_3(s)+SiO_2(s)},
\end{align}
can occur. This reaction traps carbon dioxide into 
carbonates that can accumulate before being buried by 
subduction. A very interesting property of this 
\textit{carbonate-silicate cycle}, is that is provides a powerful 
stabilizing feedback on planetary climates on geological timescales 
\cite[]{WHK81} as detailed in section~\ref{sc:carb}. On the Earth, most of the
initial CO$_2$ inventory is thought to be trapped in the 
form of carbonates in the
crust after chemical precipitation. The formation of carbonates has
also been suggested on early Mars when abundant liquid water 
seems to have flowed on the surface, possibly explaining the fate of an early
thick CO$_2$ atmosphere \cite[]{Poll:87}.
However, almost no carbonates were initially detected by the OMEGA imaging
spectrometer in spite of its high sensitivity to the spectral
signature of carbonates \cite[]{Bibr:05}.
Recently, several observations from orbiters \cite[]{Ehlm:08,Cart:12}
and landers \cite[]{Boyn:09,Morr:10} 
have revived the carbonate hypothesis
and reasserted the importance of carbon dioxide chemistry in martian climate history
\cite[]{Harv:10}.

%Indeed, because the reaction occurs in an aqueous phase 
%and is made easier by the erosion of the ground by rainfalls, 
%weathering rates increase when precipitations, and thus temperatures, 
%increase. On the other extreme, weathering is fairly limited when the 
%surface is frozen. As a result, high (low) temperatures will increase 
%(decrease) the weathering, leading to a lower (higher) greenhouse 
%effect by carbon dioxide. This feedback, which is believed to have 
%stabilized Earth climate over time, is at the very heart of the modern 
%vision of the concept of "habitable zone", the zone around a star where 
%liquid water can be stable at the surface of the planet \cite[]{KWR93}.

%However, this means that the climate itself can have a strong feedback 
%on the atmospheric composition. For example, in the absence of liquid 
%water, the capture of CO$_2$ is very inefficient. Thus, while Venus and 
%the Earth may have accreted a comparable amount of CO$_2$ during their 
%formation, it might be because Venus climate has entered a runaway 
%greenhouse state for which no liquid water could be stable at the 
%surface that the planet ended up with a thick, 93\,bar CO$_2$ 
%atmosphere \cite[]{RD70}.

\subsection{ Major classes of atmospheres } 

\label{sec:atm_class}

\begin{figure}[hbtp]
  \centering
    \includegraphics[width=14cm]{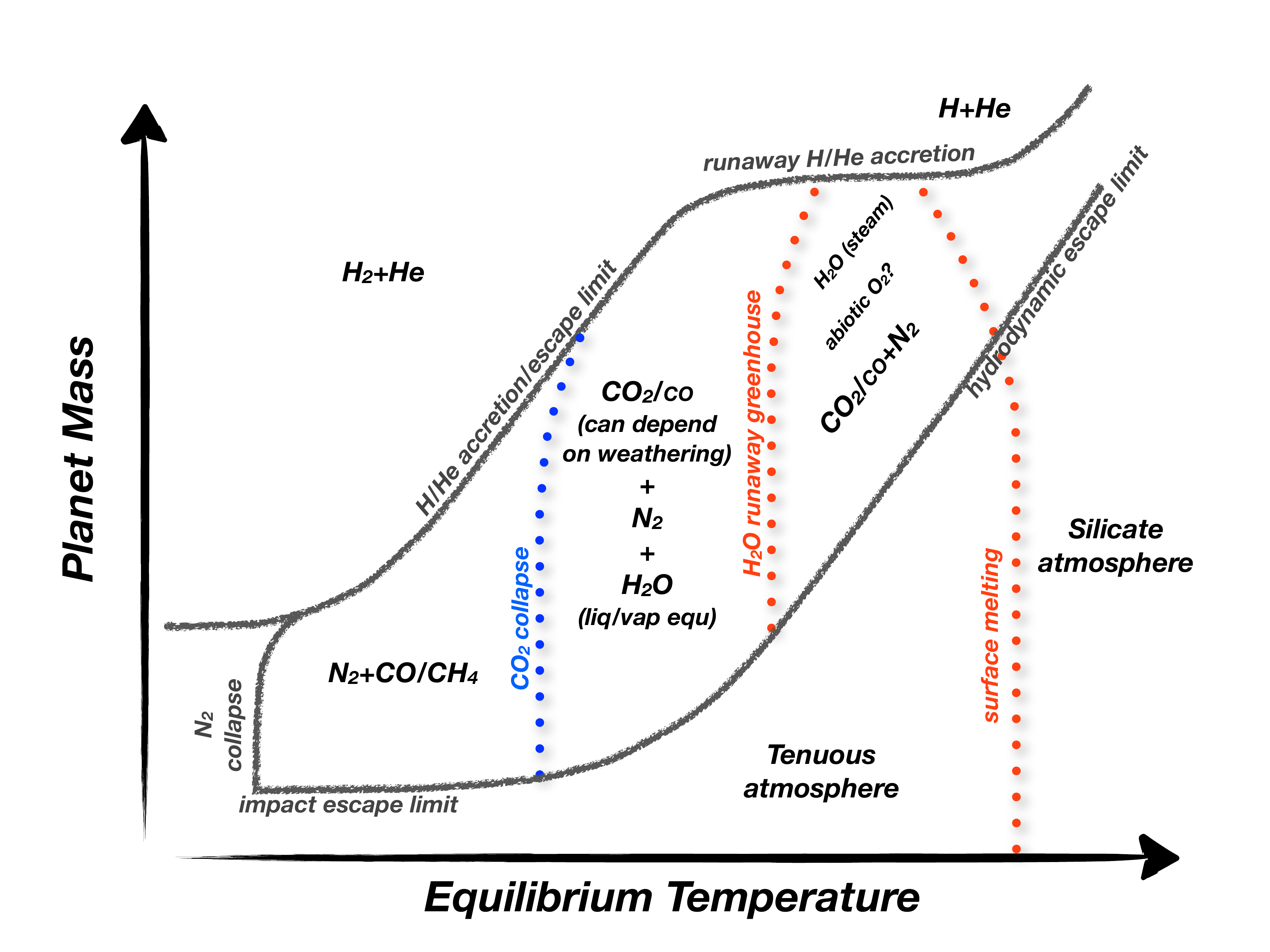}
    \caption{Schematic summary of the various class of atmospheres. 
Each line represent a transition from one regime to another, but note 
that these "transitions" are in no way hard limits. Only the expected 
dominant species are indicated, but other trace gas should of course be 
present.  }
  \label{fig:diagram}
\end{figure}

Let's make an attempt to see how the processes described above
fit together to produce the diversity of atmospheres 
that we know, or can expect. Indeed, we have seen that they
have the ability to build or get rid of an 
entire atmosphere. The key is now to identify the mechanism(s) that are 
most relevant to a given planetary environment. The result of this 
process is summarized in the diagram shown in Figure~\ref{fig:diagram} and 
discussed below.

%%%%%%%%%%%%%%%%%%%%%%%%%%%%%%%%%%%%%%%%%%%%%%%%%%%%%%%%%%%%%%%%%
%%%%%%%%%%%%%%%%%%%%%%%%%%%%%%%%%%%%%%%%%%%%%%%%%%%%%%%%%%%%%%%%%

\subsubsection{H/He dominated}

Hydrogen and helium being the lightest elements and the first to be 
accreted, they can most easily escape. The occurrence of H/He dominated 
atmospheres should thus be limited to objects more massive than the 
Earth. Indeed, in the Solar System, none of the terrestrial planetary 
body managed to accrete or keep a potential primordial H/He envelope, 
even the coldest ones which are less prone to escape.

Figure~\ref{fig:MR} suggests that a mass as low as 2$\,M_\mathrm{Earth}$ can be 
sufficient to build and keep such an atmosphere. But it also suggests 
that being more massive than that is by no means a sufficient 
condition. Indeed, some objects have a bulk density similar to the 
Earth up to 8-10$\,M_\mathrm{Earth}$. Although these high density planets 
receive a stronger stellar insolation on average, it is not clear yet 
whether this correlation stems from the fact that planets forming in
closer orbits can accrete less nebular gas \cite[]{IH12}, or from the 
fact that hotter planets exhibit higher escape rates. Note that the 
first hypothesis assumes that such close-in planets are formed in-situ.

Then, the presence of a large fraction of primordial nebular gas in the 
atmosphere of warm to cold planets above a few Earth masses should be 
fairly common. The real question will be to know the atmosphere mass 
and by how much these atmospheres will be enriched in heavy elements 
compared to the parent star? Such information will be critical to 
better understand the early stages of planet and atmosphere formation 
during the nebular phase.

%a completer, en particulier sur l'enrichissement!!!

\subsubsection{H$_2$O/CO$_2$/N$_2$ atmospheres } 

Then, if, for some reason, the planet ends up its accretion phase with 
a thin enough H/He atmosphere so that surface temperatures can be cold 
enough for the solidification of the rocky surface, 
a significant amount of H$_2$O and 
CO$_2$ should be released (envelope between gray curves in 
Figure~\ref{fig:diagram}). To understand what will happen to these volatiles, 
however, one needs to understand in which climate regime the planet 
will settle.

Saving for later the very hot temperatures for which the surface itself 
is molten, let us go through the different available regimes from the 
upper left to the lower right part of Figure~\ref{fig:diagram}.

Above a certain critical flux, the so called runaway greenhouse limit, 
the positive radiative feedback of water is so strong that the 
atmosphere warms up until surface water is vaporized 
\cite[]{Kas88}\footnote{Although this limit is not that well defined 
when the water inventory is very limited \cite[]{AAS11,LFC13}.}. In this 
case, the absence of surface water hampers CO$_2$ weathering, leaving 
most of the CO$_2$ inventory in the atmosphere.

In this case, a key question concerns the conservation of the water 
itself. Indeed, if H$_2$O is a major constituent of the atmosphere, it 
can easily be photo-dissociated high up. This produces H atoms that are 
ready to escape. Although this seemed to occur on Venus, more massive 
planets with a higher gravity to counteract escape, or objects that 
accreted more water may still possess a significant fraction of 
atmospheric water (see for example the debate on the atmospheric 
composition of GJ\,1214\,b; \citealt{MF10,DBD12})

Below the runaway greenhouse limit, water can condense at the surface. 
Except for a few planets very near the limit, water vapor should thus 
remain a trace gas in liquid/vapor equilibrium with the surface. Thus, 
the atmosphere could be dominated by species that are less abundant in 
the initial inventory but have been slowly outgassed such as N$_2$, 
among others. In such a state, CO$_2$ weathering can be efficient so 
that the amount of carbon dioxide might depend on the surface 
temperature (see section~\ref{sc:carb}).
%\cite[]{ACC12}. 
However, if water is lost due to atmospheric 
escape, especially for lower mass planets such as Mars, or hotter ones, 
CO$_2$ could build up in the atmosphere and become the dominant gas.

For colder climates, even CO$_2$ greenhouse warming is insufficient to 
prevent its condensation indefinitely. When this CO$_2$ collapse 
occurs, water of course, but also CO$_2$ itself can only be found in 
trace amounts. As is seen in the solar system (e.g. Titan), N$_2$ thus 
becomes the only stable abundant species (apart from H$_2$/He). Carbon 
compounds can be found in the form of CO or CH$_4$ depending on the 
oxidizing/reducing power of the atmosphere. This can continue until the 
triple point of nitrogen itself is reached. At that point N$_2$ ice 
albedo feedback favors a very cold climate where nitrogen is in 
condensation/sublimation equilibrium with the surface, leaving only 
thin atmospheres such as the ones found on Pluto and Triton.

\subsubsection{Photochemistry and CO}

CO$_2$ atmosphere may not be photochemically stable. In fact, 
the abundance of CO$_2$ on Mars and Venus seemed puzzling early in the space age
\cite[]{Mcel:70} 
%[McElroy and Hunten, 1970)
because CO$_2$ is
readily photodissociated. The direct three-body recombination,
CO + O + M $\rightarrow$  CO$_2$+ M, is spin-forbidden and
therefore extremely slow at atmospheric temperatures. The
solution for Mars is that photolysis of water vapor produces
OH radicals that react readily with CO to make CO$_2$; in
effect water vapor photolysis catalyses the recombination of
CO$_2$.  Could the equilibrium be reversed in favor of CO in some 
conditions? \cite{Zahn:08} showed that this may happen in thick cold (and
thus relatively dry) atmosphere, although they noted that in reality CO could
react with the surface (Fe) and be recycled as CO$_2$ by another path. Another
interesting point is that the stability question is asymmetric. 
Under plausible conditions, a significant CO
atmosphere can be converted to a CO$_2$ atmosphere quickly in 
the case of any event
(impact, volcanism) 
that may provide water vapor, whereas it takes tens to hundreds of millions of
years to convert from CO$_2$ back to CO.  
In any case, the behaviour of CO atmosphere would be somewhat 
different that CO$_2$. 
CO is a weaker greenhouse gas than CO$_2$, but it condenses at a significantly
lower temperature (or higher pressure) than CO$_2$. In very cold cases,
conversion to CO may thus prevent atmospheric collapse into 
CO$_2$ ice glaciers.

\subsubsection{the possibility of abiotic O$_2$}

Molecular oxygen O$_2$ cannot easily become a dominating species in a
planetary atmosphere because it is chemically reactive and is
not among the volatile species provided by planetesimals.
Most of the O$_2$ in Earth's present atmosphere is thought to have
been produced by biological oxygenic photosynthesis.
Nevertheless, several abiotic scenarios 
that could lead to oxygen-rich atmospheres have been suggested and
studied in detail, because the presence of O$_2$ 
and the related species O$_3$ (easier to detect) 
in the atmosphere of an exoplanet are considered
to be possible biomarker compounds \citep{Owen:80,Lege:93}

The most likely 
situation in which O$_2$ might accumulate and become a dominant species
is a runaway greenhouse
planet, like early Venus, on which large amounts of 
hydrogen escape from a hot, moist atmosphere 
(See \citealt{Lamm:11}, and reference therein). Because
the hydrogen originates from H$_2$O, oxygen is left behind. The
escape of a terrestrial ocean equivalent of hydrogen, unaccompanied
by oxygen sinks, could leave an atmosphere containing
up to 240 bars of O$_2$ \citep{Segu:07}.
Alternatively $O_2$ could be produced by
photolysis of CO$_2$ in a very dry environment, but its concentration is
then not likely to reach more than a few percents \citep{Sels:02,Segu:07}

\subsubsection{Thin silicate atmospheres}

For low mass objects and very hot objects 
(lower part of Figure~\ref{fig:diagram}), 
escape is supposed to be efficient. Bodies in this part of the diagram 
are thus expected to have no atmosphere or possibly a very teneous
exosphere. This class actually encompasses many 
Solar System bodies: Mercury, the Moon, Ganymede, Callisto, etc.
On such bodies, a teneous gaseous enveloppe can be maintained by the
release of light molecules and atoms from the surface because of the 
energetic radiation and charged particles impacting the surface
(e.g. O$_2$ on Ganymede and Europe) or the release of gases such as  
radon and helium resulting from radiative decay within the crust and
mantle (e.g. Argon and Helium on the Moon). These are not atmospheres.

An interesting cases is Io, 
which is characterized  by an intense volcanic activity 
resulting from tidal heating from friction
generated within Io's interior as it is pulled between Jupiter and the
other Galilean satellites. This activity allows for the formation of an
extremely thin and varying atmosphere consisting mainly of sulfur
dioxide (SO$_2$)

In extrasolar systems, another exotic situation can arise. Indeed, some 
planets, such as CoRoT-7\,b \cite[]{LRS09}, are so close to their host 
star that the temperatures reached on the dayside are sufficient to 
melt the rocky surface itself. As a result, some elements usually referred to 
as "refractory" become more volatile and can form a thin "silicate" 
atmosphere \cite[]{SF09,LGF11}. Depending on the composition of the 
crust, the most abundant species should be, by decreasing abundance, 
Na, K, O$_2$, O and SiO. Interestingly enough, the 
energy-redistribution effect of such an atmosphere could be limited to 
the day side of the planet as condensation occurs rapidly near the 
terminators \cite[]{CM11}. In addition, silicate clouds could form. Both 
of these effects should have a significant impact on the shape of both 
primary- and secondary-transit lightcurves, allowing us to constrain 
this scenario in the near future.

\section{On the importance of planetary rotation}

\label{sc:rotation}

\subsection{Rotation and climate}

Besides the atmospheric composition and the mean insolation, 
one of the key parameters
that determines a planetary climate is the rotation of the body (period,
obliquity). 
Rotation rate and obliquity are thought to 
influence the climate in two ways. 
On the one hand, it governs the latitudinal 
distribution of insolation as well as the seasonal and diurnal cycle. On the
other hand, modelling studies, laboratory experiments, and our experience
in the solar system show that the atmospheric circulation and transport
directly depends on the rotation rate
via the Coriolis and centrifugal 
forces. They control 
the extension of the Hadley circulation and the formation of 
extratropical jets (in the strongly rotating regime), 
the type of planetary waves, and
the tendancy of slowly rotating planets toward supperotation
(see \citealt{Show:13, Read:11pss} and references therein).

Because of the angular momentum accreted during their formation, 
most planets initially tend to rotate around their axis relatively quickly.
In the solar system, all planets which have
not been significantly influenced by the gravitation of another body rotate with
a period of about one Earth day or less (e.g. Mars, Jupiter, Saturn, Uranus,
Neptune).  However, during its existence, the rotation of a body is modified by
tidal effects resulting from gravitational forces from its parent star 
or from its satellites (or from its parent planet in the cases of satellites). 
These forces tend to cancel out the obliquity (creating
poles where almost no starlight reaches the surface) 
 and synchronize the rotation rate (possibly creating a
permanent night side).  They can thus strongly influence the climate.

\label{sc:extra}

\subsection{Tidal evolution of planetary spin}

In which cases will a planet be affected by gravitational tides ?	
When an extended, deformable body is orbited by another mass, the 
differential gravitational attraction of the latter always causes the 
primary object to be distorted. These periodic deformations 
create friction inside the deformable body which dissipates 
mechanical energy and allows angular momentum to be exchanged between 
the orbit and the spin of the two orbiting objects. In general, such tidal 
interactions eventually lead to an equilibrium state where the orbit 
is circular and the two components of the system are in a spin-orbit 
synchronized state with a zero obliquity \cite[]{Hut80}. However, 
because the tidal potential decreases as the distance between the two 
bodies to the minus sixth power, this equilibrium can take several 
dozens of billions of years to be achieved.

In a star planet system, because the angular momentum contained in 
the planetary spin is small compared to the orbital and stellar one, 
the evolution of the planetary spin (synchronization and alignment) 
is the most effective and occurs first. Indeed, the standard theory 
of equilibrium tides \cite[]{Dar1880,hut81,LCB10} predicts that a 
planet should synchronize on a timescale equal to

\begin{equation}
\tau_\mathrm{syn}=\frac{1}{3}r_\mathrm{g}^2 \frac{M_\mathrm{p}\, a^6}{G M_\mathrm{s}^2 R_\mathrm{p}^3 k_\mathrm{2,p}\,\Delta t},
\end{equation}

where $M_\mathrm{p}$ and $R_\mathrm{p}$ are the planetary mass and 
radius, $M_\mathrm{s}$ is the stellar mass, $r_\mathrm{g}$ is the 
dimensionless gyration radius ($r_\mathrm{g}^2=2/5$ for a homogeneous 
interior; \citealt{LLC11}), $G$ is the universal gravitational 
constant, $k_\mathrm{2,p}$ the tidal Love number of degree 2 that 
characterizes the elastic response of the planet, and $\Delta t$ a 
time lag that characterizes the efficiency of the tidal dissipation 
into the planet's interior (the higher $\Delta t$, the higher the 
dissipation). While the exact magnitude of the tidal dissipation in 
terrestrial planets remains difficult to assess, one can have a rough 
idea of the orders of magnitude involved by using the time lag 
derived for the Earth from the analysis of Lunar Laser Ranging 
experiments, $k_\mathrm{2,p}\,\Delta t=0.305\times629$\,s 
\cite[]{NL97}. This yields a synchronization timescale of 20\,Gyr for 
the Earth, 3\,\,Gyr for Venus and 80\,Myr for Mercury. This is 
consistent with the fact that, while the Earth has been able to keep 
a significant obliquity and a rapid rotation, both Venus and Mercury 
have a rotation axis aligned with the orbit axis and a slow rotation, 
although this rotation is not synchronous (see hereafter). The fact 
that Mercury's orbit is still eccentric confirms that tidal 
circularization proceeds on longer timescale. In many extrasolar 
systems where planets are found much closer to the central star, 
tidal synchronization and coplanarization of the planetary rotation 
is thus expected to be fairly common. In particular, rocky planets 
within or closer than the habitable zone of M and K stars are thought 
to be significantly tidally evolved. This can have a profound impact 
on the climate of these planets as it creates permanent cold traps 
for volatiles at least near the poles and possibly on 
the permanent dark side if the rotation rate is fully synchronized.

However, when a terrestrial planet has a permanent bulk mass 
distribution asymmetry or possesses a thick atmosphere, tidal 
synchronization is not the only possible spin state attainable by the 
planet. In the first case, as for Mercury, if the planet started from 
an initially rapidly rotating state, it could become trapped in 
multiple spin orbit resonances during its quick tidal spin down 
because of its eccentric orbit. This is also expected for extrasolar 
planets. Because Gl\,581\,d has a non-negligible eccentricity, it has 
a high probability of being captured in 3:2 (three rotation 
per two orbits) or higher resonance before reaching full 
synchronization \cite[]{MBE12}. In the second case, thermal tides in 
the thick atmosphere can create a torque that drives the planet out 
of the usual tidal equilibrium. This is what is thought to cause the 
slow retrograde rotation of Venus \cite[]{DI80,CL03}, and it could 
also even lead extrasolar terrestrial planets out of synchronization 
\cite[]{CLL08}. In any cases, these states all have a low obliquity, 
which would have an important impact on the climate by creating cold 
poles.

%\section{Outstanding questions and opportunities for 
%EChO}\label{sec:opportunity}

%Because in our Solar System we often have only one or two particular 
%examples of planets, if any, in each of the aforementioned classes, it 
%is difficult to quantify the location of the various transitions in 
%Figure~\ref{fig:diagram}. This is where extrasolar planets can prove an 
%invaluable asset. Indeed, if we are able to identify the main 
%constituents of tens to hundreds of exoplanets in various 
%mass/temperature regimes, we will not be looking at individual cases 
%anymore, but at populations.

%Such a global view is critical if we truly want to understand the 
%process of atmosphere formation as a whole, and how it behaves in 
%various environments. Indeed, to that purpose, we need to know what is 
%the fraction of super-Venuses compared to super-Mercuries or 
%mini-Neptunes, for example. Only a dedicated transit spectroscopy 
%mission can tackle such an issue. Interestingly, many 
%atmospheric-regime transitions occur in the high-mass/high-temperature 
%part of Figure~\ref{fig:diagram} which is exactly where observatories like 
%EChO are most sensitive. This means that, even before being able to 
%characterize an Earth-like planet in the habitable zone, we will be 
%able to characterize exotic terrestrial-planet atmospheres in key 
%regimes that are mostly unheard of in the Solar System.

\section{Which climate for a given atmosphere?}

\label{sc:clim}

\subsection{Key processes in a climate system}

Any planetary atmosphere exhibits an apparent
high level of complexity, due to a
large number of degrees of freedom, the
interaction of various scales, and the fact that atmospheres tend to
propagate many kind of waves.

\begin{figure}
\centerline{
\includegraphics[width=12cm,clip]{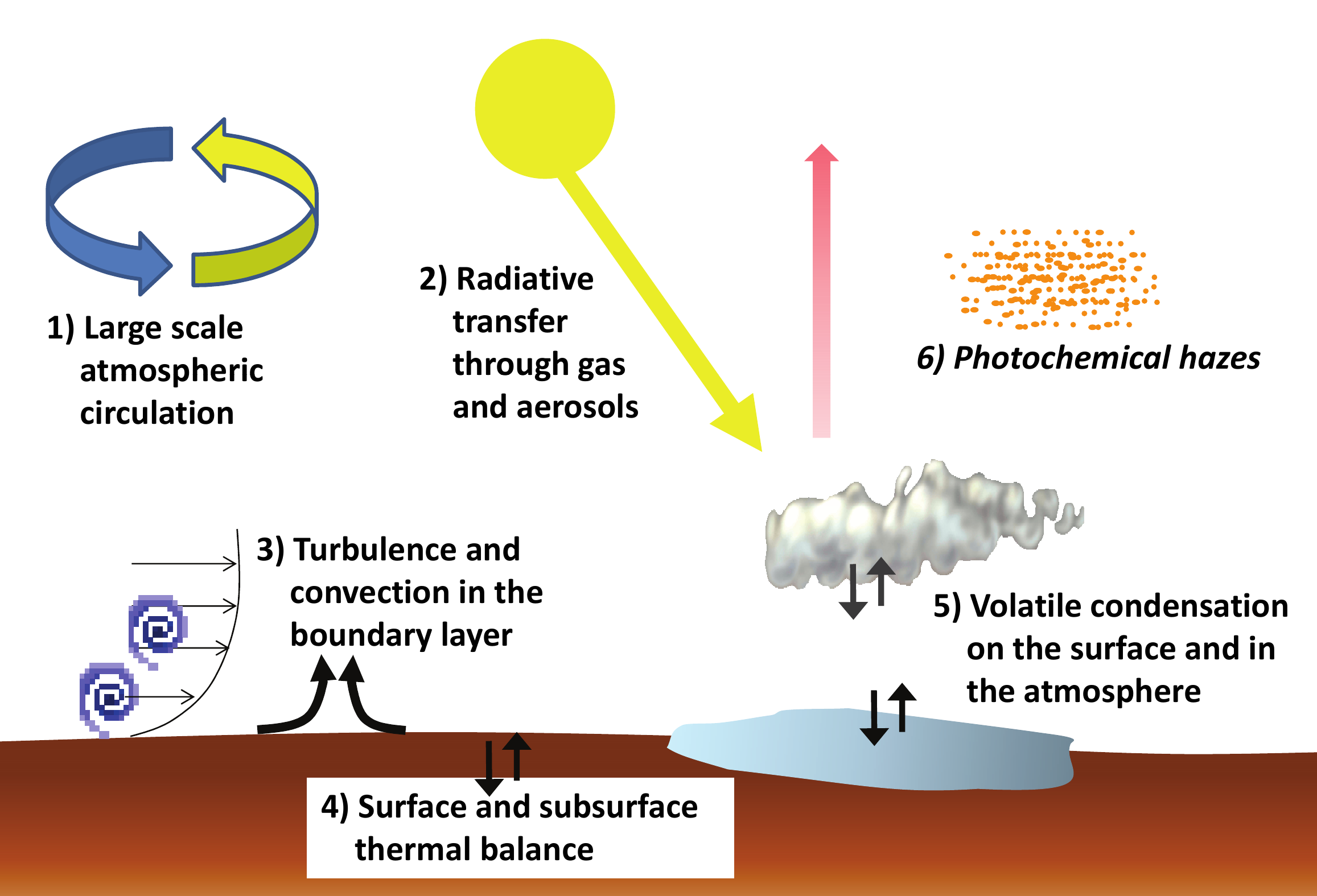}
}
\caption{
The key physical and dynamical processes which control a terrestrial 
planet climate. In practice, when modelling planetary climates,
these processes can be parametrized independently and combined to create  
a realistic planetary global climate model.
\label{fg:processes}}
\end{figure}

However the key physical and dynamical processes at work 
on a terrestrial planet
are in finite number. 
To first order, on most planets, the following coupled 
dynamical and physical processes control the climate 
(Figure~\ref{fg:processes}): 

1) Radiative transfer of stellar and thermal radiation 
through gas and aerosols. 

2) The general circulation of the atmosphere
primarily forced by the large scale, radiatively induced
temperature gradients

3) Vertical mixing and transport due to small 
scale turbulence and convection. 

4) The storage and conduction of heat in the subsurface.

5) The phase changes of volatiles on the surface and in the
atmosphere (clouds and aerosols).

6) To this list, one could add a catalog of processes which 
are only relevant in particular cases, or which play a secondary
role:  Photochemistry 
(producing aerosols, hazes or creating spatial inhomogeneities in the 
atmospheric composition), 
mineral dust lifting, oceanic transport, molecular diffusion 
and conduction (at very low pressure), etc.

Depending upon the planet's physical characteristics (orbit, size, rotation, 
host star, etc.) and of the composition of its atmosphere, 
the combination of all these processes 
can lead to a variety of 
climates that we will not try to describe here. Instead, we discuss below our
ability to simulate and predict the diversity of these terrestrial 
climates, using numerical models.

%%%%%%%%%%%%%%%%%%%%%%%%%%%%%%%%%%%%%%%%%%%%%%%%%%%%%%%%%ù
%%%%%%%%%%%%%%%%%%%%%%%%%%%%%%%%%%%%%%%%%%%%%%%%%%%%%%%%%ù

\subsection{Modelling terrestrial planetary climate}

\subsubsection{From 1D to 3D realistic models }

The processes listed above can be described with a limited number of coupled
differential equations, and it is now possible to develop numerical climate
models to predict the environment on terrestrial planets.

Until recently, a majority of studies on terrestrial exoplanets 
had been performed with simple  1D 
steady-state radiative convective models.
They can evaluate the global mean conditions on a given planet 
resulting from the radiative properties of its atmosphere and the insolation
from its star (see for instance the reference paper on habitability
by \citealt{Kast:93}). 
Such 1D models have been extremely 
useful  to explore the possible climate regimes, although
they are often not sufficient to predict the actual state of a planet, and in
particular represent the formation, distribution and radiative impact of
clouds, or	to simulate local conditions at a given time 
(for instance due to the diurnal and seasonal cycles). 
3D models are especially necessary to estimate
the poleward and/or nightside transport of energy by the atmosphere and, in
principle, the oceans. 

Exploring and understanding the atmospheric transport and the
possible  circulation regime as a function of
the planet characteristics is a research field by itself. It does not
require complete realistic climate models. For this purpose
dynamicists have used models with
a 3D hydrodynamical ``core'' designed to solve the
Navier-Stokes fluid dynamical equations in the case of a rotating
spherical envelope,  forced with a
simplified physics to represent the possible thermal gradients
(see a review in \citealt{Show:13}
as well as the recent work by
\citealt{Read:11pss,Heng:11,Edso:11})

More complete three dimensional 
numerical global climate models (``GCMs'') can be built
by combining the various components which are necessary to 
simulate the major processes listed above (Fig.~\ref{fg:processes})~: 
A 3D hydrodynamical ``core'', and, 
for each grid-point of the model, a radiative transfer
solver, a parametrization of the turbulence and convection, a subsurface thermal
model, a cloud model, etc. 
Such models have been developped (in some cases for more than twenty years)
for the telluric atmospheres in the solar system:
 the Earth (of course), Mars  Venus,
Titan, Triton and Pluto. 
The ambition behind the development
of these GCMs is high : the ultimate goal is to
build numerical simulators only based on universal
physical or chemical equations, yet able to reproduce
or predict all the available observations on a given
planet, without any ad-hoc forcing. In other words, we
aim at creating in our computers virtual planets ``behaving''
exactly like the actual planets.
In reality of course, nature is always more complex than expected,
but one can learn a lot in the process. 
In particular, a key question is now to
assess whether the GCM approach, tested in the solar system,  is
``universal'' enough to simulate the diversity of
possible climate on terrestrial exoplanets and  
accurate enough to predict the possible climate in specific cases.

Several teams are now working on the development of 3D global
climate models  designed to simulate any type of terrestrial climate,
i.e.
with any atmospheric cocktail of gases, clouds and aerosols, for any
planetary size, and around any star. For instance, at LMD, we have  
recently developed such
a tool (see e.g. \citealt{Word:11ajl,Word:13,Forg:13,Leco:13}),
by combining the necessary parametrizations listed above.
One challenge has been to develop a radiative transfer
code fast enough for 3D simulations and versatile enough to model any
atmospheric composition accurately.
For this purpose, we used the
the correlated-k distribution technique. We also included a dynamical
representation of heat transport and sea-ice formation on a potential ocean,
from \cite{Codr:12}. 

\subsubsection{What have we learned from our experience in the solar system?}

A first lesson from the modelling of the (limited) diversity of climate in the
solar system is that the same equations are often valid 
in several environments, and
that the different model components that make a climate model 
can be applied without major changes to most terrestrial planets.
Of course, the spectroscopic properties of the atmosphere, for instance, 
must be adapted in each case. Atmospheric radiative properties have been
well studied in the Earth's case for which numerous spectroscopic
databases are available. However some unknowns remain for observed
atmospheres like Mars or Venus, and
many more uncertainties affect the modelling of
theoretical exotic atmospheres not yet observed
(e.g. hot and wet atmospheres
with a high partial pressure of water vapor). 
The parametrization of large scale
dynamics, turbulent mixing, and subsurface heat conduction have been
applied to different planets without modification and 
comparisons with the available observations have not revealed major
problems. 
In some cases,  some simplifications 
that were initially done for the Earth's case 
(constant atmospheric composition, ``thin atmosphere 
approximation'') must  be questioned on other planets. For instance, 
on Venus the air specific heat varies significantly (around 40\%) with
temperature from the surface to the atmosphere above the clouds, whereas it 
was assumed to be constant in the dynamical cores derived from Earth modelling.
The consequences on the potential
temperature and dynamical core were discussed in \cite{Lebo:10}.

A second major lesson is that, by
many measures, Global Climate Models work.  They have been able
to predict the behaviour of many aspects of several climate systems
on the basis of physical equations only. Listing the success of Global Climate
models on other planets is out of the scope of this paper, 
but we could mention a few examples (with a 
biases toward our models developed at LMD).
On Mars 
assuming the right amount of dust in
the atmosphere, it has been relatively easy to
simulate the thermal structure of the atmosphere and the behaviour
of atmospheric waves such as thermal tides and baroclinic waves
\cite[]{Habe:93,Hour:95,Wils:96,Wils:97,Forg:99,Lewi:99,Ange:04}.
, to reproduce the main seasonal characteristics of the water cycle 
\cite[]{Rich:02water,Mont:04jgr}, or to predict the detailed 
behaviour of ozone \cite[]{Perr:06,Lefe:08}. On Titan, GCMs have 
anticipated the superrotating wind fields with amplitude and characteristics
comparable to observations \cite[]{Hour:95b,Newm:11}, and allowed to simulate 
and interpret the detached haze layers \citep{Rann:02,Rann:04},
 the abundance and vertical profiles of most chemical
compounds in the stratosphere, and their enrichment in the winter
polar region \cite[]{Lebo:09}, the distribution of clouds
\cite[]{Rann:06}, or the detailed thermal
structure observed by Huygens in the lowest 5~km \cite[]{Char:12}.
On Venus, the development of ``full'' GCMs ( i.e. at least coupling a 3D
dynamical core  and a realistic radiative transfer)
is more recent, but these models successfully reproduce 
the main features of the thermal structure and the superrotation of the
atmosphere \cite[]{Iked:07,Lebo:10,Lee:12,Mend:12}.

A third and even more interesting lessons is related to the ``failure''
 of planetary
Global Climate Models \citep{Forg:13}. 
When and why GCMs have not been able to predict 
the observations accurately?
Different sources of errors and challenges are listed
below:
\paragraph{Missing physical processes.} As can be expected, in many cases GCMs
fail to accurately simulate an observed phenomenon simply because a physical
process is not included in the GCM. For instance,
for many years, the thin water ice clouds present in the Martian
atmosphere had been assumed to have a limited impact on the Martian climate.
Recently, several teams have included their effect in GCM simulations
\cite[]{Made:12,Wils:11,Kahr:12,Read:11mamo}.

What they found is that not only do the clouds affect the thermal structure
locally, but that their radiative effects
could solve several long-lasting Mars climate enigmas like the pause in
baroclinic waves around winter solstice, the intensity of regional dust storms
in the northern mid latitudes, or the strength of a thermal inversion observed
above the southern winter pole.

%\nocite{Made:12,Wils:11,Kahr:11,Read:11mamo}

\paragraph{Positive feedbacks and instability.}
Another challenge is present for climate modellers when the system is very
sensitive to a parameter because of positive feedbacks. A well known example is
the albedo of snow and sea-ice on the Earth. If one tries to model the Earth
climate systems ``from scratch,
it is rapidly obvious that this model parameter must be tuned to
ensure a realistic climate at high and mid-latitude. An overestimation of the
ice albedo results in colder temperatures, more ice and snow, etc... 

\paragraph{Nonlinear behaviour and threshold effects.} An extreme version of
the model sensitivity problem is present when the climate depends on processes
which are non-linear or which depend on poorly understood physics. For instance,
the main source of variability in the Mars climate system is related
to the local, regional and sometimes global dust storms that occur on
Mars in seasons and locations that vary from year to year. This dust cycle
remains poorly understood, possibly because the lifting of dust occurs
above a local given wind threshold stress which may or may not be reached
depending on
the meteorological conditions. As a result, modelling the dust cycle and in
particular the interannual variability of global dust storms remains one of the
major challenges in planetary climatology -- not mentioning an hypothetical
ability to predict the dust storms
\citep{Newm:02a,Basu:06,Mulh:13}.
%(Newman et al. 2002, Basu et al. 2006). 
Most likely, in addition to the
threshold effect, a physical process related to the evolution of the surface
dust reservoirs is missing in the models.

\paragraph{Complex sub-grid scale processes.}
Another variant of the problems mentioned above can be directly attributed to
processes which cannot be resolved by the dynamical core, but
which play a major role in the planetary climate. Mars dust storms would be
once again a good example, but the most striking example is the representation
of sub-grid scale
clouds in the Earth GCMs. 
The parametrization of clouds
has been identified as a major source of
disagreement between models and uncertainties 
when predicting the future of our planet
(see e.g. \citealt{Dufr:08}).

\paragraph{Weak forcings, long timescales.}
While different GCMs can easily agree between themselves and with the
observations when modelling a system strongly forced by the variations of, say,
insolation,
GCM simulations naturally become model sensitive when the evolution of the
system primarily depends on a subtle balance between modeled processes. 

An
interesting case is the Venus general circulation.
the superrotation
of Venus atmosphere is the
result of a subtle equilibrium involving
 balance in the exchanges of angular
momentum between surface and atmosphere, and balance in the angular momentum
transport between the mean meridional circulation and the planetary waves,
thermal tides, and gravity waves. 
Comparative studies between Venus GCMs under identical physical
forcings \cite[]{Lee:10,Lebo:13} have recently shown
that modeling this balance is extremely 
 sensitive to the dynamical core details, to the
boundary conditions and possibly also to initial conditions.
These studies revealed that various dynamical cores, which would give very similar
results in Earth or Mars conditions, can predict very different circulation
patterns in Venus-like conditions.

\paragraph{The take-home message.}
Overall, our experience in the Solar System has shown that the 
different model components that make a climate model can be applied 
without major changes to most terrestrial planets. It has also 
revealed potential weaknesses and inaccuracies of GCMs.  Clearly, 
when modelling climate systems which are poorly observed, it is 
necessary to carefully explore the sensitivity of the modeled system 
to key parameters, in order to ``bracket'' the reality. Nevertheless 
it seems to us that when speculating about the climate regime on a 
specific detected planet and in particular its habitability, the 
primary uncertainty lies in our ability to predict and imagine the 
possible nature of its atmosphere, rather than in our capacity to 
model the climate processes for a given atmosphere. This said, that 
does not mean that it is easy to predict temperatures and the states of the
volatiles, because in many cases an accurate climate model exhibits a 
very high sensitivity to some parameters, as detailed in the next section.

\section{Climate instability and feedbacks}
\label{sc:instability}

\subsection{Runaway glaciation and runaway greenhouse effects.}

While studying the sensitivity of Earth's climate 
and the extension of the habitable zone (i.e. the range of orbits
where the climate can be suitable for surface liquid water and life), it has
been discovered that the climate of a planet with liquid water 
on its surface can be extremely sensitive 
to parameters such as
the radiative flux received from its parent star. This results from the fact
that the radiative effect of water strongly varies 
with temperature as its phase
changes, inducing strong feedbacks \citep{Kast:93}. 

For instance, slightly ``moving'' a 
planet like the Earth away from the Sun induces
a strong climate instability because of 
the process of ``runaway glaciation'': a lower solar flux
decreases the surface temperatures, and thus increases the snow and ice cover,
leading to higher surface albedos which tend to further decrease the surface
temperature \citep{Sell:69,Gera:92,Long:97}. 
The Earth would become completely frozen (and several
tens of degrees colder on average) if moved away from the Sun 
beyond a threshold distance which is highly model dependent, but probably close
to the present orbit (5 to 15\% further from the Sun). 
Furthermore, there is an hysteresis. This Earth-like planet
would remain ice-covered when set back to its initial conditions: 
Being
completely frozen is thus, in theory, another stable regime for the Earth.
Notice that this ice-snow albedo feedback is much weaker
around M-stars because water ice tends then to have a much lower albedo,
since it absorbs in the near infrared
where M-stars emit a significant fraction of their radiation
\citep{Josh:12}.

Alternatively,
when a planet with liquid water on its surface is ``moved'' toward its sun, its
surface warms, increasing the amount of water vapor in the atmosphere. 
This water vapor strongly enhances the greenhouse effect, which tends to further warm
the surface. This ``runaway greenhouse'' process can destabilize the climate
On the basis of simple 1D model calculations, Kasting (1988) found
that on an Earth-like planet around the Sun, oceans would completely 
vaporize below a threshold around
0.84 Astronomical Units (AU). He also showed that the stratosphere
would become completely saturated by water vapor at only 0.95 AU.
There, it could 
be rapidly dissociated by ultraviolet radiation, with the hydrogen
lost to space (the Earth currently keeps its water thanks to the cold-trapping
of water at the tropopause). 

In all these examples, a lot of uncertainties exist, 
especially in relation to the role of
clouds, the actual ice-snow albedo, 
the spectroscopy of water vapor, the transport of heat by the 
atmosphere and ocean, etc.
The threshold values obtained are highly model dependent. 
Nevertheless these famous cases tell us that real climates 
systems can be affected by strong instabilities which can drive 
planets subject to a similar volatiles inventory and forcing 
to completely different states. This is probably not limited to water.
At colder temperatures,
 the concept of runaway greenhouse and runaway glaciation can be extended to
CO$_2$ (which can influence the albedo or the atmospheric greenhouse effect by
condensing onto the surface or subliming), or even N$_2$ \citep{Pierr:09}. 

The uncertainties
related to the volatiles phase changes are even higher in the cases of tidally
evolved planets (see Section~\ref{sc:rotation}), 
like in the habitable zone around M-stars. Then the
permanent night side (for a planet locked in a 1:1 resonnance) 
or at least the permanently cold polar regions 
(for planet with very small obliquities) are cold traps on which 
water and other consensable atmospheric gases 
like CO$_2$ or N$_2$ 
may permanently freeze, possibly inducing an
atmospheric collapse. For a given planet, this allows, in theory,  additional
stable climate solutions (see e.g. \citealt{Josh:97,Pierr:11,Leco:13}).  
Nevertheless, It can be
noticed that on such slowly rotating bodies the atmospheric dynamics can be very
efficient to transport heat from the sunlit regions to the nightside and
therefore homogenize the temperatures and prevent atmospheric collapse 
\citep{Josh:97,Josh:03,Word:11ajl}. 
In fact the habitability of planets around an M-dwarf could
actually ``benefit'' from tidal locking. The cold trap on the
night side may allow some water to subsist well inside the inner edge of the
classical habitable zone. If a thick
ice cap can accumulate there, gravity driven ice flows and geothermal flux
could come into play to produce long-lived liquid water at the edge and/or
bottom of the ice cap \citep{Leco:13}. 
Similarly, on a cold locked ocean planet, 
the temperature at the substellar point can be much higher than 
the planetary average temperature.  
An open liquid water pool may form around the substellar point
within an otherwise frozen planet  \citep{Pierr:11}. 

\subsection{Climate stabilization and plate tectonics}
\label{sc:carb}

Another concept that must be taken into account when speculating about the
possible climates on terrestrial planets is the possibility that 
a planet can be influenced by negative feedbacks, which will ultimately 
control its atmosphere and drive it into a specific regime. 

For instance
such a scenario
is necessary to explain the long-term habitability of the Earth, which has been
able to maintain liquid on its surface throughout much of 
its existence in spite of a
varying solar luminosity and changes in its atmospheric composition which could
have led it to runaway glaciation.

Most likely, this has been possible
thanks to a long-term stabilization of the surface
temperature and CO$_2$ level due to the carbonate-silicate cycle 
\citep{Walk:81,Kast:93}. As mentioned in Section~\ref{sc:weathering}, 
on Earth, CO$_2$ is permanently  removed from the atmosphere
by the weathering of calcium and magnesium silicates in rocks and soil,
releasing various  ions, including carbon ions (HCO$_{3}^{-}$,
CO$_{3}^{2-})$. These ions are transported into the world ocean through
river or ground water runoff. There, they form carbonate and
precipitate to the seafloor to make carbonate sediments.
Ultimately, the seafloor is subducted into the mantle, where
silicates are reformed  and CO$_2$ is released
and vented back to the atmosphere by volcanos.
Assuming that weathering is an increasing function of the
mean surface temperatures (through a presumed enhanced role of the
water cycle, precipitation, runoff,  with higher temperatures)
one can see that this cycle can stabilize the climate, because the
abundance, and thus the  greenhouse effect of CO$_2$ 
increases with decreasing temperatures, and vice-versa.
This mechanism is thought to be efficient for any sea-land fraction,
although the climate stabilisation may be limited on pure waterworld
without subaerial land on which temperature-dependent weathering may
occur \citep{Abbo:12}.

On  the Earth,
the key process allowing the carbonate-silicate cycle -and more
generally the long-term recycling of atmospheric components chemically trapped
at the surface- is thus 
plate tectonics. This is a very peculiar regime induced by
the convection in the mantle, the failure of the lithosphere (the ``rigid layer''
forming the plates that include the crust and the uppermost mantle). 
and surface cooling. 

How likely is the existence of plate tectonics elsewhere?
In the solar system, Earth plate tectonics is unique and its
origin not well understood. Other terrestrial planets or satellites are
characterized by a single ``rigid lid'' plate surrounding the planet, and this may
be the default regime on extrasolar terrestrial planets. 
On
planets smaller than the Earth (e.g., Mars), the rapid interior cooling
corresponds to a weak convection stress and a thick lithosphere, and no plate
tectonics is expected to be maintained in the long term. On larger planets
(i.e., ``Super-Earths''), available studies have reached very different 
views \citep{Vale:07,Vanh:11,Onei:07,Stei:13}. 
It is possible that in super-Earth
the large planetary radius acts to decrease the ratio of convective
stresses to lithospheric resistance \citep{Onei:07} and that 
the very high internal pressure
increases the viscosity near the core-mantle boundary, resulting in a highly
``sluggish'' convection regime in the lower mantles of those planets which may
reduce the ability of plate tectonics \citep{Stei:11,Stam:12}.

What these studies highlight is the possibility that the
Earth may be very ``lucky'' to be in an exact size range (within a few percents)
that allows for plate tectonics. Furthermore, Venus, which is about the size of
the Earth but does not exhibit plate tectonics, shows that the Earth case may be
rare, and that many factors control the phenomenon. On Venus, for instance, it
is thought that the mantle is drier than on Earth, and that consequently it is
more viscous and the lithosphere thicker \citep{Stei:11,Nimm:98}. Similar
considerations led \cite{Kore:10} to conclude that the likelihood of plate
tectonics is also controlled largely by the presence of surface water. Plate
tectonics may also strongly depend on the history and the evolution of the
planet. Using their state of the art model of coupled mantle convection and
planetary tectonics, \cite{Lena:12} found that multiple tectonic
modes could exist for equivalent planetary parameter values, depending on the
specific geologic and climatic history. 

\section{Conclusions}

Based on the examples observed in the solar system, and on the available
observations of exoplanets, we can
expect a huge diversity among exoplanetary climates. In the absence of direct
observations, one can only speculate on the various possible cases. According to
models, climate should primarily depend on 
1) The atmospheric composition and mass
and the volatiles inventory (including water);
2) The incident stellar flux (i.e. the distance to the parent star); 
3) The tidal evolution of the planetary spin (which also depend of the 
distance to the star for a planet). 

In theory, the atmospheric composition and mass depends on complex processes
which are difficult to model: origins of volatiles, atmospheric escape,
geochemistry, long-term photochemistry. 
Some physical constraints
exists which can help us to speculate on what may or 
may not exist, depending on
the planet size, its final distance from its star 
and the star type and its
activity (Figure~\ref{fig:diagram}). Nevertheless 
the diversity of atmospheric composition remains a field for which new
observations are necessary. 
Once a type of atmosphere can be assumed, theoretical
3D climate studies, which benefit from our experience in modelling terrestrial
atmospheres in the solar system, should allow us to estimate the range of
possibilities, and in particular estimate if 
liquid water can be stable on the surface of these bodies and if they can be
habitable.
Whatever the accuracy of the models, predicting the actual climate
regime on a specific planet will remain challenging because climate systems are
affected by strong positive feedbacks (runaway glaciation and 
the runaway greenhouse
effect) which can drive planets subject to a similar
volatiles inventory and forcing and to completely different states.

We can hope that in the future it will be possible to learn more
about exoplanetary  atmospheres thanks to telescopic observations and
spectroscopy. An important step will be achieved in the next decade by space
telescopes like the James Webb Space Telescope (JWST) or the proposed 
ECHO mission \citep{Tine:12}, as well as by  Earth-based telescopic observations using new generation
telescope like the European Extremely Large Telescope. These projects will
notably be able to perform atmospheric spectroscopy on exoplanets transiting in
front of their star as seen from the Earth. Characterizing atmospheres
of terrestrial planets in or near the habitable zone will remain challenging.
Furthermore, the  number of observable planets at a suitable distance will
probably be very low. Nevertheless, well before the time when we will be able to
detect and characterize a truly habitable planet, the first observations of
terrestrial exoplanet atmospheres, whatever they show, will allow us to make a
major progress in our understanding of planetary climates and 
therefore in our estimation of the likelihood of life 
elsewhere in the universe.    

\section*{Acknowledgment}

F. Forget and J. Leconte wish to thanks another 
Forget and Leconte team who, through their achievements in the 
Davies Cup on November 29 - December 1 1991, 
has inspired them toward teamwork, friendship and fun.

%%%%%%%%%% Insert bibliography here %%%%%%%%%%%%%%
% \bibliography{jeremy,new,newfred}
% \bibliographystyle{apalike}

\end{document}